\documentclass[9pt]{article}
\usepackage{graphicx}
\usepackage{subcaption}
\usepackage{dcolumn}
\usepackage{booktabs}
\usepackage{changepage}
\usepackage[utf8]{inputenc}	
\usepackage[sorting=none, citestyle=numeric-comp]{biblatex} 
\usepackage{float}
\usepackage[hidelinks]{hyperref}
\usepackage[left=1cm, right=1cm, top=1cm, bottom=1cm, bindingoffset=0cm]{geometry}
\usepackage{setspace}
\usepackage{mathtools}
\usepackage{authblk}
\renewcommand{\figurename}{Fig.}

\title{Position sensitive resonant Schottky cavities for heavy ion storage rings}
\author[1,2]{D. Dmytriiev}
\author[1]{M. S. Sanjari}
\author[1,2]{Yu. A. Litvinov}
\author[1,3]{Th. Stöhlker}
\affil[1]{GSI Helmholtzzentrum für Schwerionenforschung, 64291 Darmstadt, Germany}
\affil[2]{Universität Heidelberg, Germany}
\affil[3]{Helmholtz-Institut Jena, Germany}
\date{}
\begin{document}
	\maketitle
\section{Introduction}
	Studying the rapid neutron capture process (r-process) in stellar environments, that leads to the creation of elements heavier than 56-Fe, remains one of the fundamental questions of modern physics and therefore an active field of research within nuclear astrophysics \cite{RevModPhys.29.547}. Apart from other key measurables like neutron capture cross section and decay lifetimes, nuclear masses are of outmost importance for pinpointing the r-process using theoretical and experimental approaches. Exotic nuclides which participate in the r-process due to their low production yield and short half-life can be efficiently investigated  in storage rings \cite{bosch_nuclear_2013, wu_performance_2013}. In such facilities non-destructive methods of particle detection are often used for in-flight measurements based on frequency analysis \cite{sanjari_resonant_2013}. Due to the low signal level the detectors should be very sensitive and fast because of short lifetime of the particles. Resonant Schottky cavity pickups fulfill such requirements \cite{nolden_fast_2011}. Apart from their applications in the measurements of beam parameters, they can be used in non-destructive in-ring decay studies of radioactive ion beams \cite{sanjari_resonant_2013}. In addition, position sensitive Schottky pick-up cavities can enhance precision in the isochronous mass measurement technique. The goal of this work is to design such a position sensitive resonant Schottky cavity pickup based on theoretical calculations and simulations. 	
\section{Methods}
We used a right-handed coordinate system where Z-axis was aimed in the same direction as particle momentum vector. It was suggested to model a cavity with an elliptical geometry to use a monopole mode of the electromagnetic (EM) field inside for establishing a position of a particle along the X-axis and also to decrease influence of the vertical position of the beam on the output signal \cite{sanjari_IBIS_2014, sanjari_conceptual_2015}. 
In order to normalize the signal from the position sensitive cavity, one has to attach a cylindrical cavity to a beam pipe to have a reference signal from it. Signals from the cylindrical cavity will provide information about mass over charge ratio whereas signal from the elliptical one will provide information about mass over charge ratio and particle position. Using both of these signals one can normalize signal from the elliptical cavity in order to get information of a particle position regardless of its mass over charge ratio. 
\section{Results}
\subsection{Modeling of the position sensitive resonant Schottky cavity}
\begin{figure}[H]
  		\centering
  		\begin{subfigure}[b]{0.48\linewidth}
  			\centering\
    		\includegraphics[width=\linewidth]{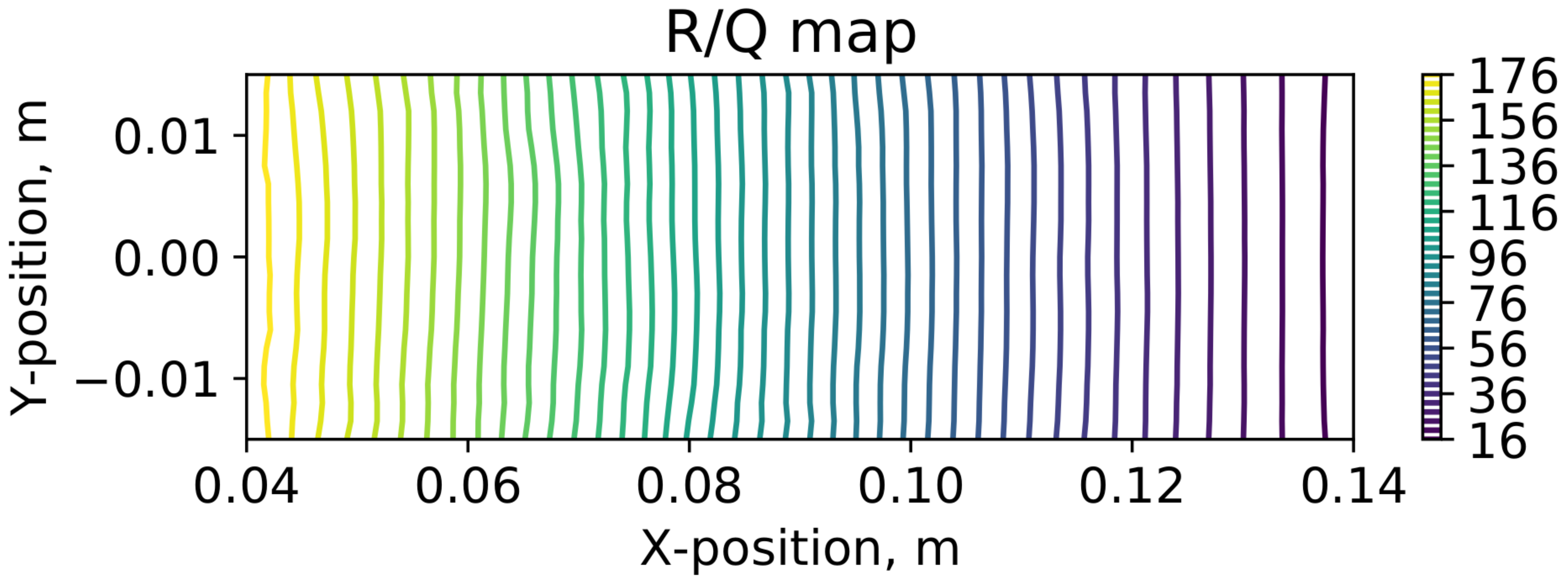}
    		\caption{Blended connection.}
  		\end{subfigure}
  		\begin{subfigure}[b]{0.48\linewidth}
  			\centering\
    		\includegraphics[width=\linewidth]{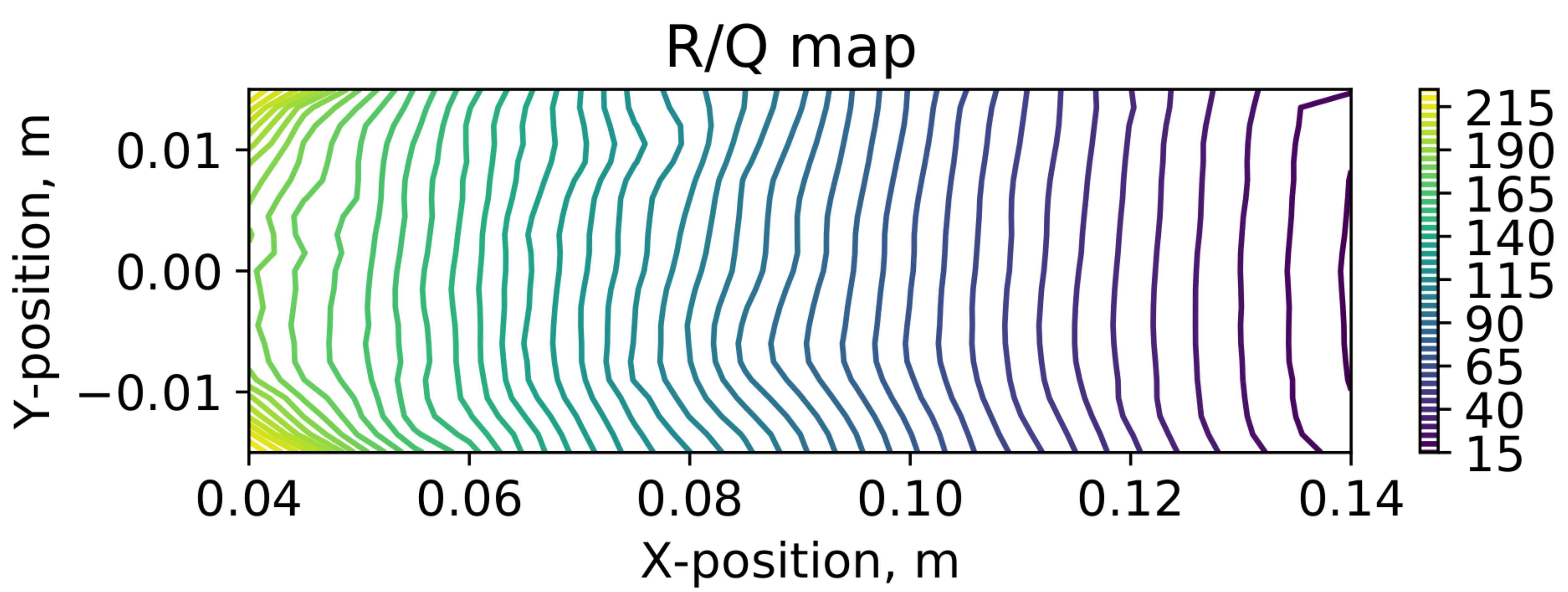}
    		\caption{Sharp-edge connection.}
  		\end{subfigure}
  		\caption{Comparison of R/Q maps with and without blending}
  		\label{fig:Blend}
	\end{figure}

\begin{figure}[H]
  		\centering
			\includegraphics[scale=0.9]{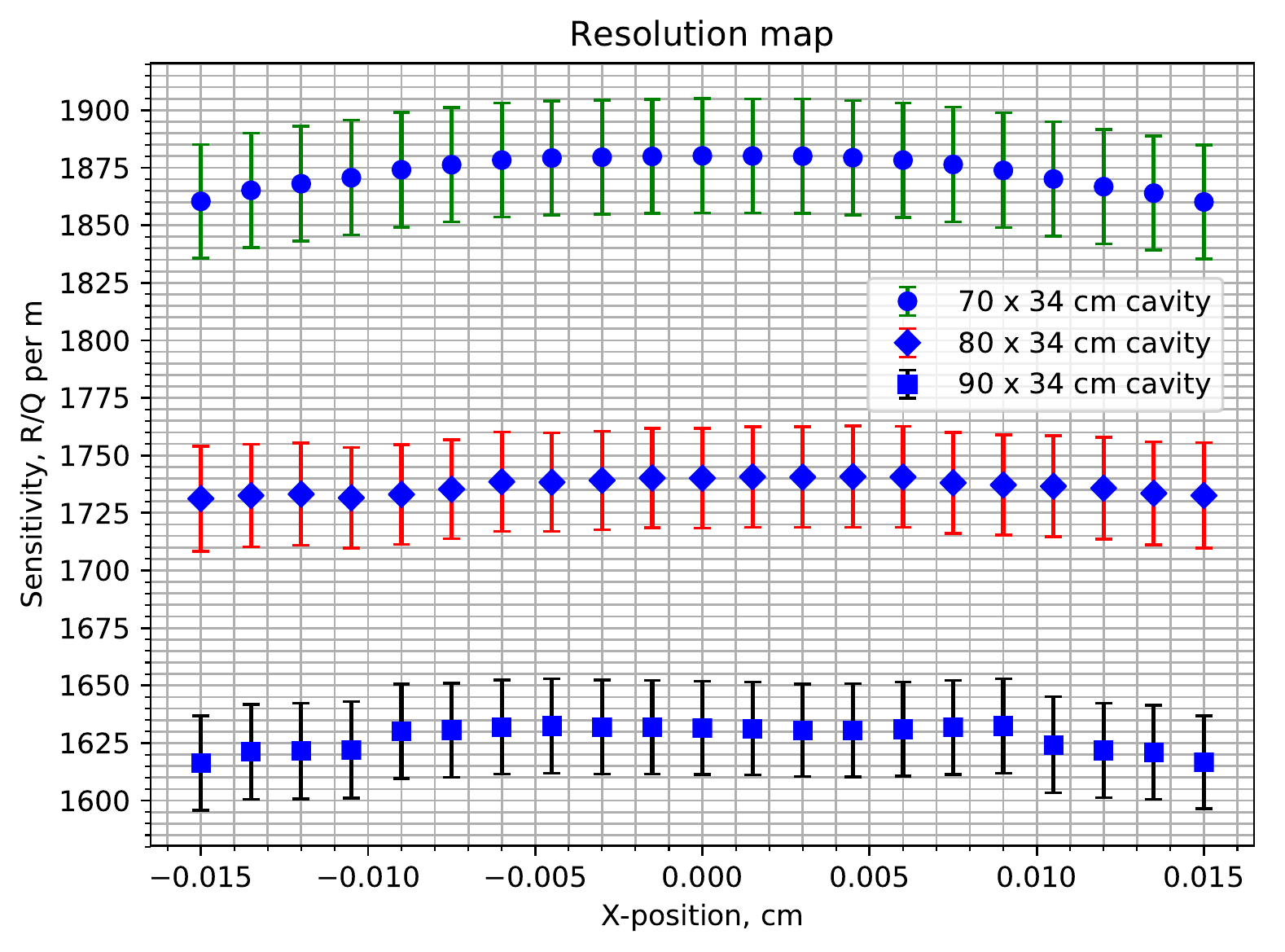}
  		\caption{Resolution maps for different cavity sizes}
  		\label{fig:Resmaps}
	\end{figure}

Using the CST Studio Suite \cite{noauthor_cst_2018} we analyzed dependency of the EM-field distribution on the geometry of the cavity. The intensity of the interaction between the beam and a cavity can be described using R/Q formalizm which is well described in \cite{m._s._sanjari_resonant_2013}. As a result of simulation modeling of the connection between beam pipe and cavity with sharp edges leads to the large distortions of the EM field inside the cavity. On the other hand blending has to avoid large radii otherwise it will affect on the cavity shape \cite{chen_intensity-sensitive_2016}. Different blending radii were simulated, as a result, we found that a blending with radius of 2 cm will be the best solution. Results of simulations are given on \figurename~\ref{fig:Blend}
The cavity has to be less than 1 m high to make transportation and mounting easier. This and size of the beam pipe were only two restrictions which we used as starting points. Of course, for a storage ring one has to calculate size of the cavity according to the beam dispersion and available space in a place where cavity is supposed to be installed.	
One of the most important parameters of a detector is its resolution. The idea of the resolution map is to show how large the horizontal position resolution is for every vertical position of the beam. To create one point at that map we took all R/Q data along certain y-coordinate inside the beam pipe, fitted it linearly and obtained the value of the slope and its error. The slope indicates the resolution in R/Q units per meter whereas the slope error describes deviations of the field gradient from the linear fit. \figurename~\ref{fig:Resmaps} shows the slope and its error for every vertical position to give the overall understanding of the cavity resolution at any point inside the beam pipe. To choose a certain geometry of the cavity we decided to take care of resolution to resolution error ratio at the point y=0 (horizontal symmetry line of the cavity). It was decided to choose the geometry such that this ratio will be the largest. Resolution to resolution error ratio was chosen as a value for comparison of different geometries because it shows how large the error given by the cavity is. There is a trade-off between gradient of the field which defines position sensitivity of the cavity and error which is caused by non-linearity in the field gradient. If we have a good sensitivity of the cavity but a large error of establishing the beam position we will be able to measure the position only with a resolution which is larger than error regardless of the gradient of the field and this can not be compensated by using external circuits. Moreover, all elements of the external circuits will add their own errors. So we need to use that size of the cavity which will add the smallest possible error to the signal. The best geometry 80 $\times$ 34 cm. This parameters of the cavity will allow us to reach the largest possible resolution to resolution error ratio which is 39.85. Some examples of the R/Q maps are given in \figurename~\ref{fig:Resmaps} 
\subsection{Distance between cavities}
\begin{figure}[H]
  		\centering
  		\begin{subfigure}[c]{0.35\linewidth}
  			\centering
    		\includegraphics[width=\linewidth]{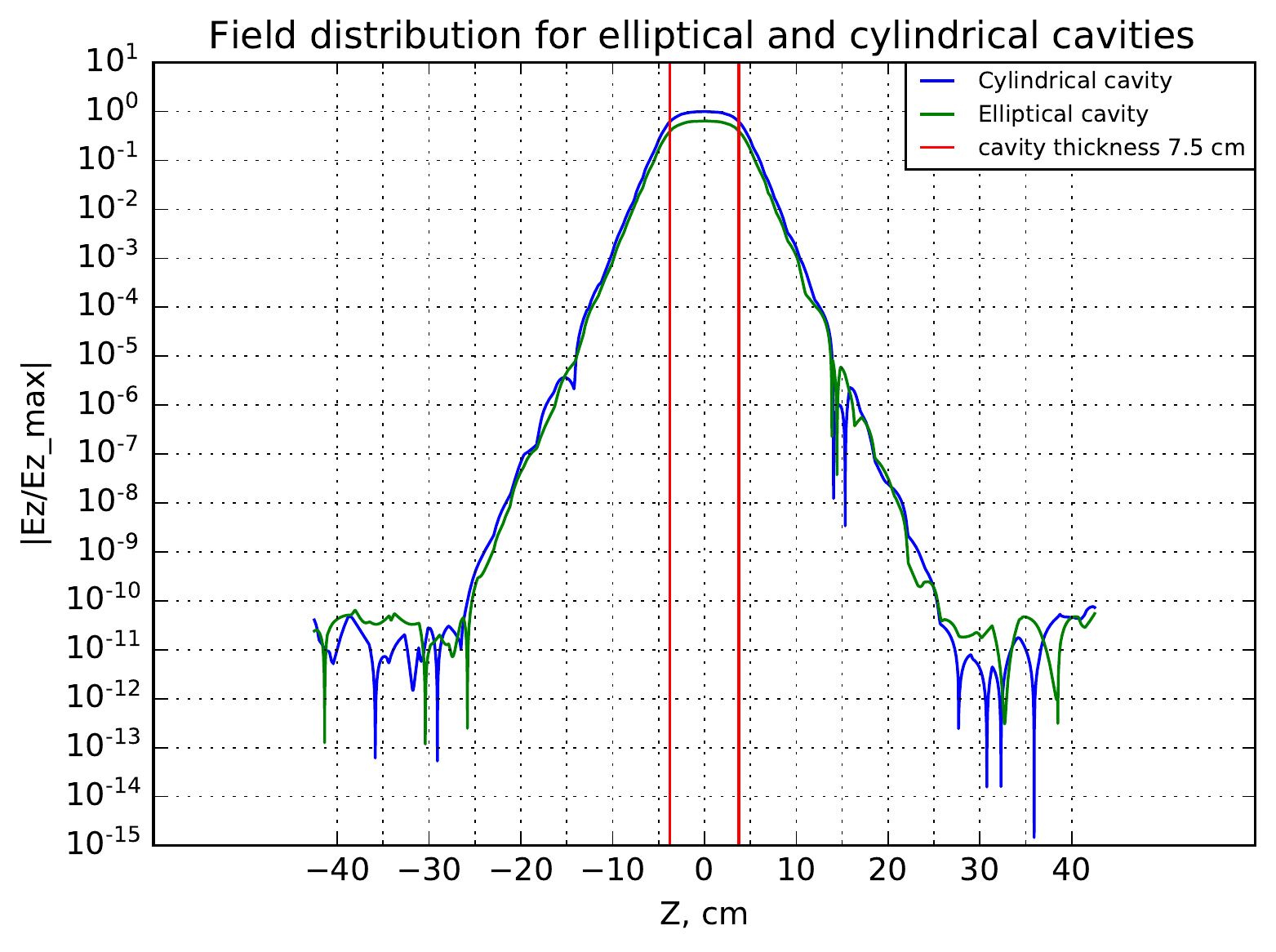}
    		\caption{Longitudinal field distribution along the center of the beam pipe}
  		\end{subfigure}
  		\begin{subfigure}[c]{0.35\linewidth}
			\centering			
			\includegraphics[scale=0.45]{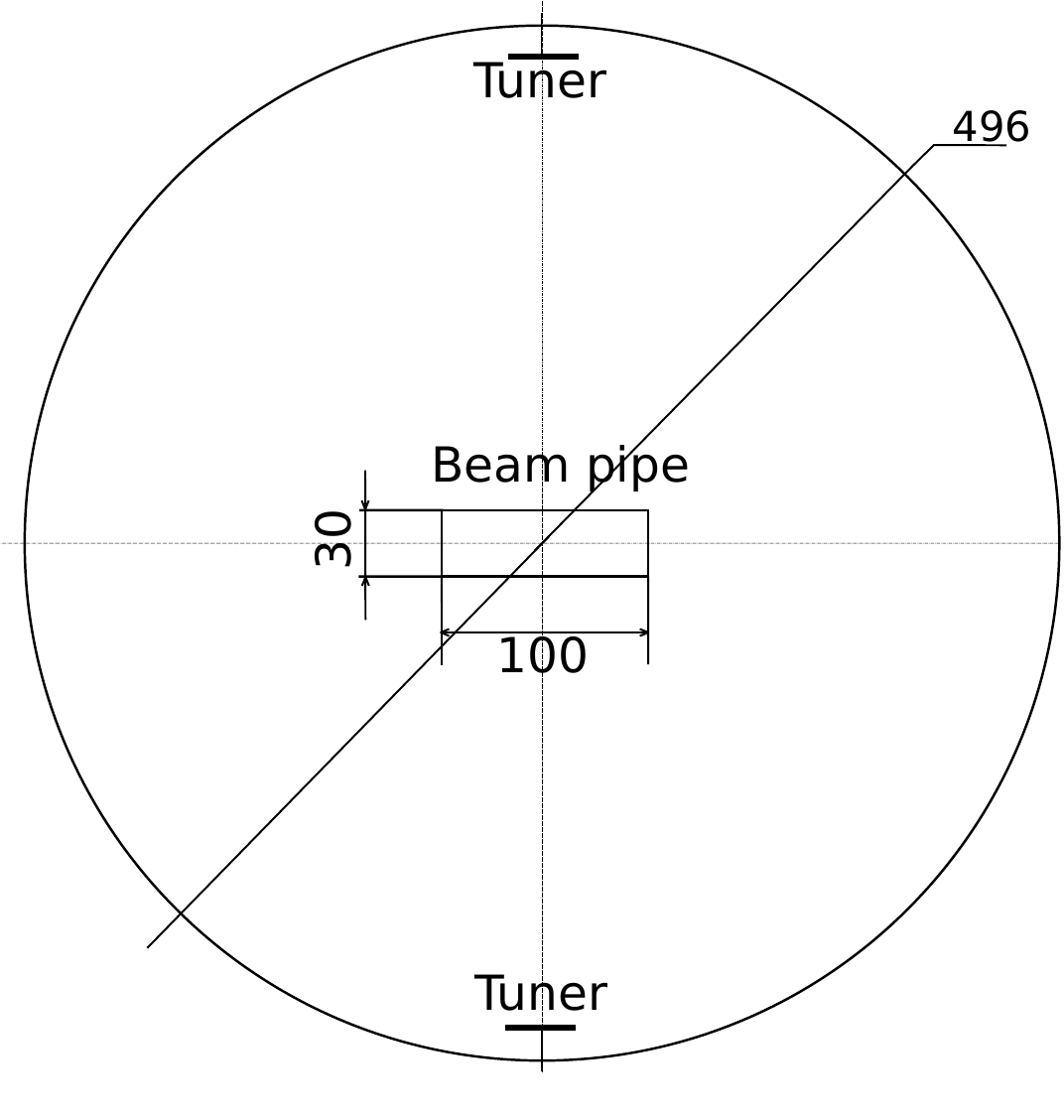}
			\caption{Sketch of the cylindrical cavity}
  		\end{subfigure}
  		\begin{subfigure}[b]{0.4\linewidth}
			\centering			
			\includegraphics[scale=0.5]{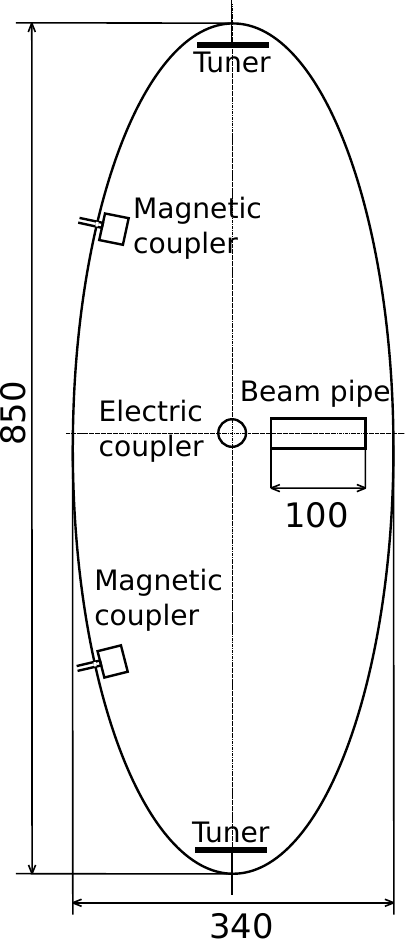}
			\caption{Sketch of the cylindrical cavity}
  		\end{subfigure}
  		\begin{subfigure}[b]{0.4\linewidth}
			\centering			
			\includegraphics[scale=0.5]{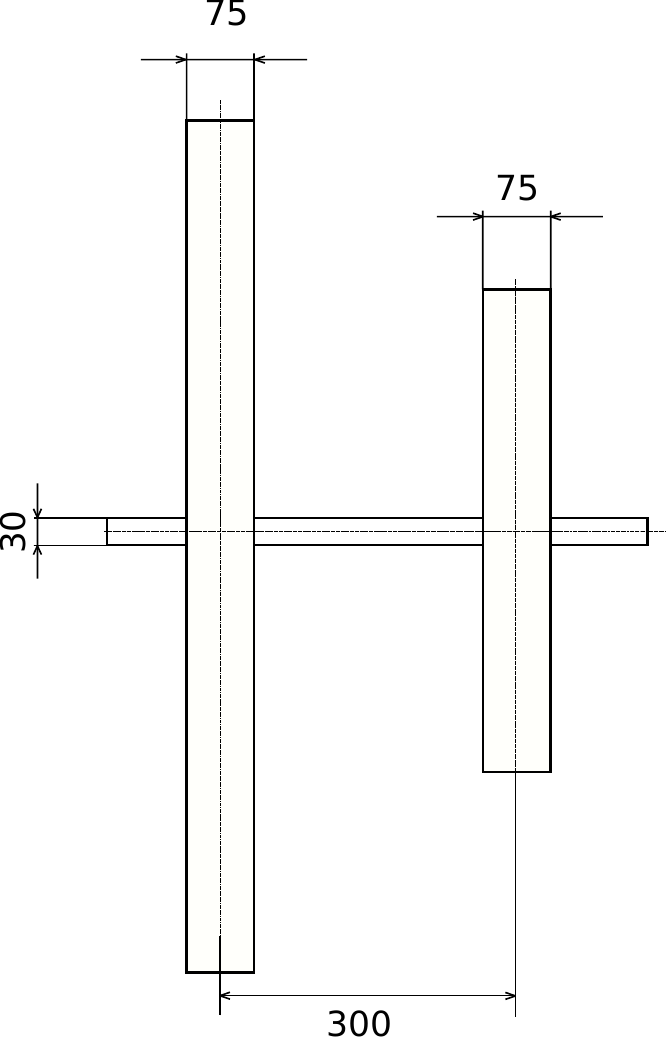}
			\caption{Sketch of the system of two cavities}
  		\end{subfigure}
  		\caption{Simulation of the EM field intensity in the system of two cavities, for (b),(c) and (d) all dimensions are in mm}
  		\label{fig:TwoCavsPlot}
	\end{figure}

Cavities have to be situated in sequence. There are two restrictions on cavities positions. First, cavities have to be as close as possible to each other in order to neglect the change of a particle position during flight through the system of both cavities. Second, cavity is not a closed resonator but has a beam pipe, so excited EM field inside the cavity will be extended along the beam pipe further than thickness of the cavity. This means that cavities have to be far enough from each other in order to avoid the influence of an extension of the field from the cylindrical cavity on a field inside the elliptical one. In addition to it construction and installation of cavities and their elements should be as simple as possible, which means, for example, that there has to be enough space for the standard mechanical parts which will be used. 
Some high frequency waves from higher order modes can propagate through the beam pipe, above the cut-off frequency. Which for rectangular waveguide with defined geometry can be defined from the formula (\ref{cutoff}):
\begin{equation}
 f_c = \frac{c}{2a}.
 \label{cutoff}
\end{equation}
Here $c$ is speed of light in the medium, $a$ is a larger side of the rectangular waveguide, $f_c$ is a cutoff frequency any wave with frequency lower than cutoff frequency can not propagate through the given waveguide. As far as we are interested in lower order modes, such as the monopole mode we can neglect the waves which flow through the beam pipe because their frequency is several times higher than frequency of the monopole mode which we want to use for position measurements. Additionally their energy is also much smaller. For the beam pipe with outer width 100 mm and wall thickness 2.5 mm this formula gives cutoff frequency which is given by the formula (\ref{cutoff}) as $\omega_0 = 1538$ MHz. 
In order to estimate the optimal distance between elliptical and cylindrical cavities field distribution of the cavities along the center of the beam pipe was investigated. Both cavities were simulated separately, but the model remains the same beam pipe with cylindrical and elliptical cavity on it. All sizes were constant during all simulations. The profile of Z component of the EM field long the Z-axis in the middle of the beam pipe was taken and "0" point for each plot is the center of the corresponding cavity. The results are shown in \figurename~\ref{fig:TwoCavsPlot}. All dimensions are given in millimeters.
According to \figurename~\ref{fig:TwoCavsPlot}  we decide that if the field intensity of the cylindrical cavity reduces 4 to 6 orders of magnitude along the longitudinal symmetry axis of the beam pipe one can neglect its influence on the resolution of the position sensitive cavity. So field distribution error caused by the interaction between EM fields of the cylindrical and elliptical cavities on the distance between cavities centers more than 30 cm is negligible.       
\section{Conclusion}
During this work  simulations for proving the general idea of using the monopole mode inside the elliptical cavity have been done. A proof of concept and first measurements were performed in \cite{m._s._sanjari_resonant_2013, chen_x._non-interceptive_2015}. As a result of the simulation in the  current work the sizes of the elliptical cavity were chosen according to the construction and accelerator parameters requirements. Sizes of the elliptical cavity are 80 $\times$ 34 cm. Resonant frequency is around 513 MHz. Radius of the cylindrical cavity which will be placed in front of elliptical one is 22.8 cm because with this size it will have first resonant frequency also around 513 MHz. In addition to it cross-talking effect between two consequent cavities and its influence on the resolution was investigated. If the distance between cavities will be 30 cm or more, the intensity of the field from the cylindrical cavity will be 5 orders of magnitude less than in the cylindrical cavity. 
\section{Acknowledgements}
	This project has received funding from the European Research Council (ERC) under the European Union’s Horizon 2020 research and innovation programme (grant agreement No 682841 "ASTRUm"). D. D. acknowledges the support by HGS-HIRe.

\printbibliography
\end{document}